\documentclass[aps,prb,twocolumn,amsmath,amssymb,floatfix]{revtex4-1}
\usepackage{tabularx}
\usepackage{bm}
\usepackage{euscript}
\usepackage{graphicx}
\usepackage{color}
\usepackage{amsfonts}
\usepackage{exscale}
\usepackage{amsbsy}
\usepackage{subfigure}
\usepackage{textcomp}

\begin{document}

\title{Quantum critical metals in $d=3+1$} \author{R. Mahajan$^1$,
  D. M. Ramirez$^1$, S. Kachru$^{1,2}$ and S. Raghu$^{1,2}$}
\affiliation{$^1$Department of Physics, Stanford University, Stanford,
  California 94305, USA} \affiliation{$^2$SLAC National Accelerator
  Laboratory, Menlo Park, CA 94025, USA}

\date{\today}

\begin{abstract}
We study the problem of disorder-free metals in the vicinity of a quantum critical point in $d=3+1$ dimensions.  We begin with perturbation theory in the `Yukawa' coupling between the electrons and undamped bosons (order parameter fluctuations) and show that the perturbation expansion breaks down below energy scales where the bosons get substantially Landau damped.  Above this scale however, we find a regime in  which low-energy fermions obtain an imaginary self-energy that varies linearly with frequency, realizing the `marginal Fermi liquid' phenomenology\cite{Varma}.  
We  discuss a large N theory in which the marginal Fermi liquid behavior is enhanced while the role of Landau damping is suppressed, and show that quasiparticles obtain a decay rate parametrically larger than their energy.

\end{abstract}

\maketitle

\section{Introduction}
Some of the most  challenging open issues in condensed matter physics involve continuous quantum phase transitions  in metals.  In the vicinity of these transitions, the fluctuations of order parameter fields can drastically alter the properties of the electrons in the metal.  Indeed, experimental measurements of thermodyamic and transport quantities  in such systems often mark a departure from  the conventional Landau Fermi liquid behavior\cite{Sachdev, Lohneysen2007}.  The focus of this paper will be on an imporant class of quantum phase transitions in metals known as Pomeranchuk instabilities, which involves order parameters that preserve lattice translation symmetry.  
Examples  include transitions to phases with ferromagnetism and nematic order.  In the case of a continuous Pomeranchuk transition, the order parameter field condenses as zero momentum and its fluctuations couple to fermions everywhere on the Fermi surface.  
There is growing experimental evidence suggesting that such transitions occur in several families of highly correlated materials\cite{Fradkin2010}.

The standard paradigm for understanding  zero temperature phase transitions in metals was introduced by Hertz\cite{Hertz1976} and  Millis\cite{Millis1993}.  Their approach involves integrating out all fermionic modes,  even those  that lie on the Fermi surface.  The effective theory, involving only the bosonic order parameter fields, is then analyzed using the  renormalization group.  In metals, this procedure is dangerous: integrating out gapless fermion modes  will produce non-local, and even {\it singular} interactions among the bosons\cite{Abanov2004}.  A more systematic approach would be to keep both the bosons and the fermions in the low energy effective theory, which one obtains by integrating out only high energy excitations.   However, due to the vastly different kinematic structure of the fermions and bosons in dimension  greater than one - the low energy fermions comprise a surface, whereas the low energy bosons are in the vicinity of a point in momentum space - there are  difficulties involved in constructing such an effective theory.

A commonly adopted view has been that the anomalous power laws present in non-Fermi liquids
cannot be understood without a controlled theory of the deep infrared behavior of the boson-fermion theory\cite{Nayak1994, Polchinski1994, Chakravarty1995, Lee2009,Metlitski2010, Mross2010}.
However, at finite temperature or frequency scales, a metal in the vicinity of a quantum critical point exhibits {\it crossovers}, not transitions, from Fermi liquid to non-Fermi liquid scaling behavior.  There is the chance, therefore, that starting from a weakly-coupled metal, one can understand the intermediate asymptotic behavior by treating the coupling between the fermions and bosons either perturbatively or in various large $N$ limits.  Clearly, theories of this sort will break down below a certain energy scale reflecting the non-analyticities associated either with the quantum critical point itself, or perhaps more interestingly, with instabilities that preclude the observation of a `naked' quantum critical point.  These energy scales can be made arbitrarily low in the theories to be discussed below. Above such scales, there are no complications associated with either a perturbative or large N description of the non-Fermi liquid scaling laws of the metal.

We carry out this procedure in $d=3+1$ dimensions, where the coupling between electrons and order parameter fluctuations is dimensionless.  We find a wide range of energy scales where the electrons obtain an imaginary self-energy that varies linearly with frequency, signifying `marginal Fermi liquid' behavior \cite{Varma}.    In $d=3+1$, one also obtains marginal Fermi liquid behavior  from RPA theories starting with  Landau damped bosons in the bare action\cite{Holstein,Oganesyan2001,Podolsky2009}.  By contrast, here, we stress that the marginal Fermi liquid arises in systematic perturbative or large N theories involving undamped bosons, with quite different dynamic and thermodynamic properties.

Our calculation is carried out in the limit where the boson bandwidth is considerably smaller than the fermion bandwidth.  In microscopic derivations of quantum critical behavior based on single-band theories, the boson and fermion bandwidths are typically comparable to one another.  However, in multi-band systems, such as the heavy fermion systems, the bandwidth associated with order parameter fluctuations can be small compared to the Fermi energy.

\section{Effective action}
Let $\psi_{\sigma}$ denote a fermion field with spin $\sigma = \uparrow, \downarrow$, and $\phi$ be a real scalar boson field corresponding to the  order parameter.  The  low energy effective action consists of decoupled actions for the fermion and boson fields, and a Yukawa coupling between fermions and bosons:
\begin{eqnarray}
\label{action}
\mathcal S &=& \int dt \int d^{d-1} x \  \mathcal  L = S_{\psi} + S_{\phi} + S_{\psi-\phi}  \nonumber \\
\mathcal L_{\psi} &=&   \bar \psi_{\sigma} \left[ i\partial_t -\epsilon(-i \nabla)  \right] \psi_{ \sigma} + \lambda_{\psi} \bar \psi_{\sigma} \bar \psi_{\sigma'} \psi_{\sigma'} \psi_{\sigma} \nonumber  \\
\mathcal L_{\phi} &=&  \frac{1}{2}\left(\partial_t \phi \right)^2 - \frac{c^2}{2} \left( \vec \nabla \phi \right)^2 -\frac{1}{2} m_B^2   \phi^2   - \frac{ \lambda_{\phi}}{4 !} \phi^4  \nonumber \\
\mathcal S_{\psi-\phi}&=& \int \frac{d^{d-1}k d \omega}{(2 \pi)^d} g_{\sigma \sigma'}(k,q) \bar \psi_{\bm k + \bm q, \sigma} \psi_{\bm k, \sigma'} \phi_{\bm q}
\end{eqnarray}
where repeated spin indices are summed.   In the second line above, $\mathcal L_{\psi}$ describes  a Landau Fermi liquid, with dispersion $\epsilon(\bm k)$ relative to the Fermi level, and with weak, short-range repulsive interactions $\lambda_{\psi}$.  We will consider a quadratic dispersion of the form $\epsilon(\bm k) = (k^2-k_F^2)/2m_F$, where $m_F$ is the  fermion band mass.   $\mathcal L_{\phi}$ describes an interacting scalar Bose field with mass $m_B$ (physically, the mass $m_B$ corresponds to the inverse correlation length and vanishes at the quantum critical point) and velocity $c$.  

The Yukawa  coupling between the fermions and  bosons in the last line of Eq. \ref{action} is most naturally written in momentum-space.  The quantity $g(k,q)$ is a  coupling function that depends in general on both the fermion momentum $\bm k$ and the momentum transfer $\bm q$.  In the low energy limit, it has a Taylor expansion of the form
\begin{equation}
\label{yukawa_expansion}
g(k,q) = g_0(\bm k_F,0) + g_1 k + g_2 q + \cdots
\end{equation}
where we have suppreseed spin indices here for clarity.  
The first term above, $g_0(\bm k_F, 0)$,  corresponds to scattering of fermions near the Fermi surface by bosons imparting  nearly zero momentum transfer.  However, the function can have non-trivial $\bm k_F$ dependence: {\it i.e.} the scattering process can vary with the solid angle along the Fermi surface.  In a rotationally-invariant system, this coupling can be labelled by the angular momentum obtained by integrating over the solid-angle.  Each angular momentum component corresponds to a distinct broken symmetry and couples to a distinct order parameter field; therefore, generically, only one angular momentum component of the Yukawa coupling, which labels the type of Pomeranchuk instability, needs to be retained in the theory.  For instance, the angular momentum zero corresponds to a ferromagnetic instability whereas angular momentum two corresponds to a nematic transition.  

The spin-dependence of the Yukawa coupling also is determined by the nature of the Pomeranchuk transition.  As is well known, Pomeranchuk instabilities occur in spin-symmetric and antisymmetric channels (the latter spontaneously break spin-rotation and time-reversal invariance).   In the case of a spin-symmetric Pomeranchuk instability, the Yukawa coupling function $g_{\sigma \sigma'} \propto \delta_{\sigma \sigma'}$ whereas for  Pomeranchuk instabilities that are antisymmetric in spins, $\phi$ corresponds to a  vector order parameter and $g_{\sigma \sigma'} $ is proportional to a Pauli matrix.  In all cases, the Yukawa coupling function respects a symmetry under which both $\phi$ and the fermion bilinear that couples to $\phi$ change sign.  

In what follows, we shall consider the angular momentum zero component (and refer to it as a constant $g$); this would correspond to a ferromagnetic quantum critical point.  We have checked, however, that our main conclusions for the power laws associated with the fermion and boson self-energies do not depend on the particular angular momentum channel.  The second and third terms in Eq. \ref{yukawa_expansion} allow for the dependence on the magnitude of the fermion momenta 
as well as the magnitude of the momentum transfer imparted by the boson respectively; they are both formally irrelevant under power counting and we shall ignore them.  

We make a final comment before proceeding.  In $d=3+1$ certain Pomeranchuk transitions are guaranteed to be first order due to the presence of cubic invariants.  Examples include the Ising nematic transition, where the Yukawa coupling has angular momentum $\ell = 2$ and couples to the fermion quadrupole density.  However, when the Yukawa coupling involves the spin of the fermions, such cubic invariants are forbidden by time-reversal invariance.  Moreover, even the Ising nematic theory can be continuous, if we imagine that the Fermi surface is not perfectly spherically symmetric, but instead has a slight amount of eccentricity.  For even an arbitrarily small eccentricity, cubic invariants are forbidden and the Ising nematic theory can be continuous.  Moreover, the slight eccentricities hardly affect the analysis in what follows.  Therefore, we shall assume that cubic invariants are forbidden and that the theories to be considered remain continuous\footnote{Non-analytic corrections to Fermi liquid theory can lead to fluctuation-induced first-order transitions, as noted in Refs.\cite{Belitz2005, Belitz2012}.  However, these effects are suppressed in the large-N limit considered here.}.  With these caveats dispelled, we now proceed with the main line of analysis.  

\section{Discussion of various asymptotic regimes}
\subsection{Perturbation theory}
In this section, we begin with perturbation theory in the Yukawa coupling $g$.  We follow a standard procedure in obtaining a  perturbation expansion in $g$ in terms of one-particle irreducible (1PI) diagrams.  Such a perturbation expansion will cease to converge below a certain energy scale as is true even for ordinary Fermi liquids; to uncover the scale where perturbation theory breaks down, we consider a cutoff $\Lambda_{IR}$ on the energy of the external lines of each diagram.  When the resulting perturbative corrections are smooth in $\Lambda_{IR}$, the cutoff can be set to zero.  However, when the corrections behave in a singular or non-analytic fashion in the cutoff, they signify the development of new physics, where the feedback between the bosons and fermions begins to alter the low energy behavior of the system.   As we discuss, the highest energy scale where non-perturbative corrections begin to dominate will be set by the scale at which the bosons get Landau damped.

\begin{figure}[t]
\begin{minipage}{0.5\linewidth}
\includegraphics[width=\linewidth]{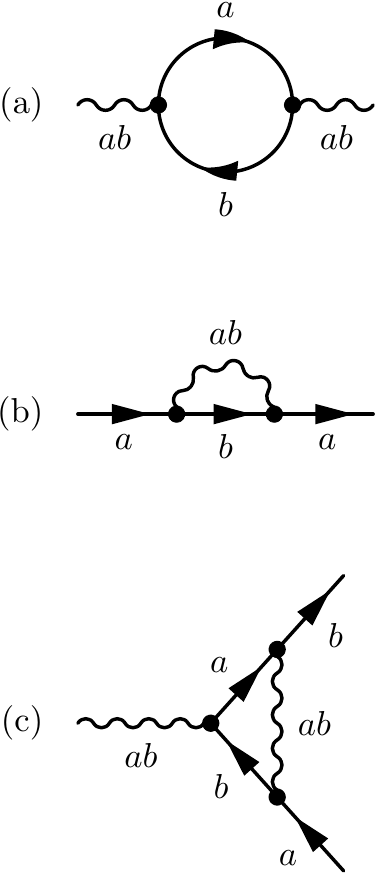}
\end{minipage}
\caption{The three one-loop Feynman diagrams relevant in our analysis.  $a$ and $b$ are flavor indices in the large $N$ version of the theory.  These indices can be ignored in section IIIA. }
\label{diagrams}
\end{figure}

The three leading one-loop Feynman diagrams are shown in Fig. \ref{diagrams}.  The first corresponds to the boson self-energy and has several effects.  Firstly, it generates smooth $\mathcal{O}(g^2)$ corrections to the boson mass and velocity which are negligible in the small $g$ limit, and in which $\Lambda_{IR}$ can be set to zero.  More importantly, it generates a non-analytic self-energy correction to the boson action, representing the overdamping of the boson from fermions near the Fermi surface.  We ignore perturbative corrections to $c$ and $m_B$ and focus on the Landau damping term:
\begin{equation}
\label{landaudamping}
\Pi(q, q_0) = i g^2\frac{m_F^2}{2 \pi} \frac{q_0}{q}
\end{equation}
This correction occurs in the limit where $q \rightarrow 0, q_0 \rightarrow 0$ with fixed $x \equiv q_0/q , \vert x \vert < 1$\cite{Doniach}.
 In Hertz's theory, this correction enters the  bare boson kinetic term.   By contrast, in a perturbative treatment, the overdamping is a correction that occurs at $\mathcal{O}(g^2)$ and is subdominant compared to the bare boson kinetic term.  However, the breakdown of perturbation theory from these diagrams occurs below an energy scale where the self-energy above is of the order of the bare inverse boson propagator:
 \begin{equation}
 g^2\frac{m_F^2}{2 \pi} x \frac{1}{\omega^2_{Ld}} = 1, \ \omega_{Ld} \sim g m_F
 \end{equation}
where  we have made use of the fact that $x \sim \mathcal{O}(1)$.  Above this scale, Landau damping is a negligible perturbative correction and the theory consists of undamped bosons.

The fermion self-energy (Fig. \ref{diagrams}(b))
\begin{equation}
\Sigma(p) = i g^2 \int \frac{d^4 k}{(2 \pi)^4} D(p-k) G(k)
\end{equation}
is evaluated in explicit detail below, where
\begin{eqnarray}
D^{-1}(q) &=& q_0^2 - c^2q^2 - m_B^2 + i \delta \nonumber \\
G^{-1}(q) &=& q_0 - \epsilon(\bm q) + i \eta {\rm sgn}\left[ \epsilon(\bm q) \right]
\end{eqnarray}
are the inverses of the bare boson and fermion propagators respectively, and both $\eta, \delta$ are positive infinitesimals.  In the analytic treatment that follows, we make two approximations.  First, we set the boson velocity to be much less than the Fermi velocity, $c \ll k_F/m_F$.  In this case, the effective ``Debye frequency", $\omega_D = 2 c k_F \ll E_F$.  Then, the boson imparts large momentum transfers, but low energy transfer.  Secondly, we neglect the variation of the density of states away from the Fermi energy, treating the density of states as a constant: $\rho(\epsilon) \simeq \rho  = m_F k_F/(2 \pi^2)$.  With these approximations, the measure of the momentum integrals is approximated by \cite{Engelsberg1963}
\begin{equation}
\int \frac{d^3 k}{(2 \pi)^3} \simeq  \frac{\rho }{(2 \pi)^2 k_F^2}  \int_0^{2 k_F}  q d q \int_{- \infty}^{\infty} d \epsilon
\end{equation}
where $ \bm q = \bm p - \bm k$.  Under these approximations, the self-energy
\begin{widetext}
\begin{eqnarray}
\Sigma(p) &=&
\frac{i g^2}{(2 \pi)^3 v_F} \int_{- \infty}^{\infty} d k_0 \int_0^{2 k_F} \frac{q d q }{(p_0-k_0)^2- c^2 q^2 - m_B^2 + i \delta} \int_{- \infty}^{\infty} \frac{d \epsilon}{k_0 - \epsilon + i \eta {\rm sgn}(\epsilon)} \nonumber \\
{\rm Im} \Sigma(p) &=&  -\frac{\pi^2 g^2}{(2 \pi)^3 v_F} \int_{-\infty}^{\infty} d k_0 {\rm sgn}(k_0) \int_0^{2 k_F} d q q \delta\left[ (p_0-k_0)^2-c^2q^2 - m_B^2 \right]
\end{eqnarray}
\end{widetext}
In the first line above,
the elementary integral over $\epsilon$  resulted in $- i \pi {\rm sgn}(k_0)$, and in the last line, we made use of the identity
\begin{equation}
\frac{1}{a + i \eta} = \mathcal{P}\left[ \frac{1}{a} \right] - i \pi \delta(a)
\end{equation}
Upon evaluating the remaining integrals, one finds
\begin{widetext}
\begin{equation}
{\rm Im} \Sigma(p_0) =
\begin{cases}
0, &  \text{ if $0 < p_0 < m_B$} \\
-\frac{g^2}{8 \pi v_F c^2} \left[ p_0 - m_B \right], & \text{if $m_B < p_0 < \sqrt{m_B^2 +  \omega_D^2}$} \\
-\frac{g^2}{8 \pi v_F c^2} \left[ \sqrt{m_B^2 +  \omega_D^2} - m_B \right], & \text{if $p_0 > \sqrt{m_B^2 + \omega_D^2}$}
\end{cases}
\end{equation}
\end{widetext}
In particular, when the boson  is tuned to criticality, $m_B \rightarrow 0$, one finds that for frequencies $p_0 < \omega_D$, the imaginary part of the electron self-energy is a {\it linear function of frequency}:
\begin{equation}
{\rm Im } \Sigma(p_0) = -\frac{g^2}{8 \pi v_F c^2} p_0
\end{equation}
and for frequencies larger than $\omega_D$, the fermions obtain a constant imaginary self-energy, as is the case for metals coupled to optical phonons.  This is the key result of this section.  From the Kramers-Kronig relations, the real part of the electron self energy is
\begin{equation}
{\rm Re} \Sigma(p_0) = \frac{g^2}{8 \pi v_F c^2} p_0 \log{\frac{E_F}{p_0}}
\end{equation}
which suggests that the perturbation theory breakdown due to the fermion self-energy correction occurs at an exponentially small scale $ \omega \sim E_F \exp{\left[-1/g^2\right]}$.
Since this is far lower than $\omega_{Ld}$, we shall neglect its effect here.

In the limit where the boson velocity approaches zero, the problem is reminiscent of that of metals coupled to optical phonons.  From the expression above, in the case where   $m_B \ne 0, c\rightarrow 0$ (which implies $\omega_D =2c k_F \rightarrow 0$) the imaginary part of the electron self-energy is a constant for frequencies above the boson mass and zero otherwise:
\begin{equation}
{\rm Im } \Sigma(p_0) = -\frac{g^2m_F k_F}{4 \pi m_B}  = -\frac{ \pi g^2 \rho}{2 m_B}, \ p_0 > m_B,
\end{equation}
This expression was first obtained by Engelsberg and Schrieffer in Ref. \onlinecite{Engelsberg1963}

The  third diagram, which corresponds to the vertex correction produces logarithmic corrections when the boson imparts zero momentum transfer.  These corrections also signify a breakdown in perturbation theory at an exponentially smaller scale than that set by the Landau damping.

To summarize, when the metal is coupled to nearly critical bosons in $d=3+1$ and when the Yukawa coupling between the fermions and bosons is treated in systematic perturbation theory, one finds a non-Fermi liquid electron self-energy that is reminiscent of the marginal Fermi liquid \cite{Varma}.  The highest energy scale at which the perturbation theory breaks down is at a scale $\omega_{Ld} = gE_F$ below which the overdamping of the bosons due to the particle-hole continuum become significant.  This scale is parametrically small compared to E$_F$ in the weak-coupling limit; above this scale the non-Fermi liquid self-energy can dominate the Fermi liquid self-energy due to momentum relaxation via fermion self-interactions
\begin{equation}
\Sigma_{FL} \sim \lambda_{\psi}^2 \omega^2.
\end{equation}
  However, in perturbation theory, the decay rate of a quasiparticle $\Gamma \sim g^2 \omega$ will always remain small compared to its energy $E \sim \omega$.  In the next section, we show in the appropriate large N limit, this is not the case and the quasiparticle becomes ill-defined.

\subsection{Large $N$ theory }
The perturbation theory presented  above showed that the overdamping of the boson represents the most important physical process that causes the theory to break down.  This overdamping is greatly enhanced in the usual large $N$ theories of metals near quantum critical points,  in which there are a large number $N$ of fermion flavors, while the number of bosons is kept at unity.  In such a theory, Landau damping plays the dominant role, as is intuitively clear from the fact that there simply are far more fermionic degrees of freedom present to damp the boson.  More formally, since the large $N$ limit is taken with $g^2 N$ held fixed, the boson self-energy is order unity, whereas the fermion self-energy is a correction at order $1/N$.  The vertex correction $\delta g/g$ is also a $1/N$ correction and can be neglected in this theory.

We are led to ask, therefore, whether there is an alternative large $N$ description of the system in which there are far more bosons than there are fermions.  In this case, the roles would be reversed - there would be many more decay channels for the fermions in the presence of a large number of bosonic variables, which would enhance the marginal Fermi liquid behavior.  Moreover, Landau damping would be a relatively muted effect, since fewer fermions are present relative to the number of bosons.  A large $N$ theory with this structure is obtained by allowing for $N$ fermion flavors transforming in the fundamental representation of $U(N)$ while describing the boson variables as an $N\times N$ matrix that transforms in the adjoint representation.  Specifically, we define a fermion spinor and boson matrix fields
 \begin{equation}
\psi_{a}, \phi_{ab}, \ a,b=1,\cdots N
\end{equation}
and analyze the behavior of the system in the  $N \rightarrow \infty, g \rightarrow 0$ limit with $g^2N \equiv g_0^2 $ held fixed.  

In $d=3+1$, the boson quartic coupling $\lambda_{\phi}$ is dangerously irrelevant at the UV fixed point, where the dynamic critical exponent of bosons is unity.   This fact plays an important role in leading to new non-Fermi liquid fixed point below $d=3+1$\cite{Fitzpatrick}.  However, in the large N limit (taken by holding  $\lambda_{\phi}N$ fixed), the fermion self-energy is independent of  $\lambda_{\phi}$.  The effect of non-zero $\lambda_{\phi}$ on the boson propagator is (i) to shift the critical point and (ii) to produce an anomalous dimension for the boson.  We shall ignore the first, since we always tune to the physical quantum critical point; the second effect occurs as a $1/N$ correction and can also be ignored in the leading large N limit.  More systematic analysis will be presented elsewhere\cite{Fitzpatrick2}.     
\begin{figure}[t]
\begin{minipage}{0.7\linewidth}
\includegraphics[width=\linewidth]{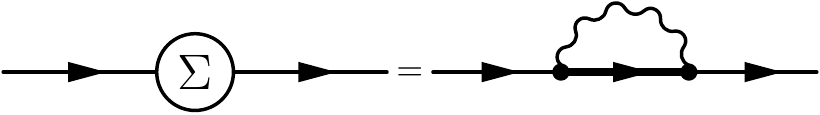}
\end{minipage}
\caption{In the large N theory in which boson fields transform in the adjoint representation of $U(N)$ (see text), the leading order electron self-energy is obtained by summing over all planar diagrams.  When boson self-interactions are absent, only the ``rainbow" diagrams above contribute to the self-energy.   }
\label{rainbow}
\end{figure}

The three one-loop diagrams that enter this theory are again the ones in Fig. \ref{diagrams}.  The boson self-energy, Fig. \ref{diagrams}(a) is a $1/N$ correction in this theory, which can simply be seen as follows.  When the external legs are fixed with a boson $\phi_{ab}$, flavor indices of the fermion lines are automatically specified in the internal loop.  The diagram therefore produces a Landau damping correction
\begin{equation}
\Pi(q, q_0) = i \frac{g_0^2}{N} \frac{m_F^2}{2 \pi} \frac{q_0}{q}
\end{equation}
and Landau damping therefore sets in below an energy scale $\omega_{Ld} = g_0 m_F/ \sqrt{N}$.  The vertex correction is similarly suppressed, and again, $\delta g/g \sim 1/N$ in this theory.  Singularities associated with the vertex correction at zero momentum transfer produce logarithmic corrections that are utterly negligible in the large $N$ limit.    By contrast, the fermion self-energy is $\mathcal{O}(1)$ as can be seen from the fact that the fermion flavor index in the internal loop can take $N$ distinct values (Fig. \ref{diagrams}b).  Therefore at criticality, $m_B=0$, the marginal Fermi liquid self-energy is obtained and describes the behavior of the system for frequencies greater than $\omega_{Ld}$.  Thus, in the present large $N$ formulation in which far more bosonic degrees of freedom are present relative to fermionic modes, marginal Fermi liquid behavior is enhanced and operates down to lower energy scales, since the effects of Landau damping are $N$-suppressed.

The large $N$ action is solved for arbitrary $g_0$  by summing over all planar diagrams\cite{Coleman}.  When boson self-interactions are negligible, the planar diagrams are  ``rainbow" diagrams for the electron self-energy (Fig. \ref{rainbow}).  The self-energy in this limit is obtained  by solving the integral equation
\begin{equation}
\Sigma(p) = i g_0^2   \int \frac{d^4 k}{(2 \pi)^4} D(p-k) G(k)
\end{equation}
for $\Sigma$, where in the right-hand side above, $G(k)$ includes the self-energy correction.  Thus, the sum over rainbows corresponds to an implicit integral equation for the electron self-energy.   We have solved the integral equation above, in the absence of boson self-interactions, $\lambda_{\phi} = 0$, with the ansatz of a momentum-independent self-energy.   We have found that, with these restrictions,  the self-energy remains proportional to $g_0^2$, since higher order rainbow graphs contain pairs of fermion propagators with the same momentum and frequency: integration over their momenta causes the contribution to vanish by Cauchy's theorem since both poles occur in the same half-plane.  This also occurs in large N QCD: for the very same reason, the quark self-energy is also proportional to $g_0^2$ when self-interactions among the gauge bosons are also neglected\cite{Hooft1973}.  Thus, the marginal Fermi liquid form of the electron self energy is the leading expression in the large N limit:
\begin{equation}
\Sigma(p)  = \frac{g_0^2}{8 \pi v_F c^2} \left[ p_0 \log{\frac{E_F}{p_0}} - i p_0 \right].
\end{equation}
Since $g_0$ is of order unity, for energy scales above the N-suppressed $\omega_{Ld} \sim g_0^2/\sqrt{N}$, a marginal Fermi liquid results in which the quasiparticle decay rate $\sim g_0^2 \omega $ can be parametrically larger than its energy.  We also remark that when the boson mass $m_B$ is finite, {\it i.e.} the system is tuned away from criticality, Fermi liquid behavior resumes at low energies.  Thus the present large N theory adequately describes a crossover from a Fermi liquid regime away from criticality to a non-Fermi liquid regime with incoherent quasiparticles as $m_B \rightarrow 0$.
\section{Discussion}

In this paper, we have discussed a simple theory in which a crossover from Fermi liquid to marginal Fermi liquid behavior occurs in the electron self-energy.  The existence of a controlled perturbative regime with marginal Fermi liquid self-energy in the simple quantum field theory of a critical boson coupled to a Fermi surface opens several attractive questions for further research.

The perturbative description of the non-Fermi liquid breaks down at a parametrically small energy scale $\omega_{Ld}$.  However, both in real quantum critical metals and in these field theories, various instabilities can arise and condense ${\it before}$ one reaches extremely low energies or temperatures.
It would be very interesting to study the emergence of superconductivity directly from the non-Fermi liquid scaling regime in this or related models.

One may also wish to find precise fixed points of the renormalization group which could represent zero-temperature non-Fermi liquid critical points or phases.  While straight perturbation theory of the sort used here does not suffice to find a controlled fixed point, it is quite plausible that a Wilsonian renormalization group treatment of a similar theory, in the $\epsilon$-expansion, could uncover controlled fixed points in $3-\epsilon$ dimensions.  This program was carried out, with interesting results, for the Landau damped boson coupled to a Fermi surface in Ref. \onlinecite{Chakravarty1995}.  The adjoint large $N$ theory described here may also be applicable in $d=2+1$.  Recently, the prevailing large $N$ theories in $d=2+1$ were shown to be ill defined even at $N=\infty$ in Ref. \onlinecite{Lee2009}.  In the alternate large $N$ theory considered here, however, the scale $\omega_{Ld}$ is again suppressed; therefore, the analysis of a system of {\it undamped} bosons in the large $N$ limit may prove to be a tractable theory, with logarithmic quasiparticle decay rates.  We shall present the results of this analysis elsewhere.

We note also the interesting similarities and differences between the theory described here, and some of the
non-Fermi liquids which have been discovered via the techniques of holography in string theory
\cite{Faulkner}.  There, a crucial insight is that coupling a Fermi surface to a `locally critical' quantum field theory at large $N$, can produce non-Fermi liquid scalings in the self-energy of the fermion
\cite{semi-holographic} (with model-dependent scaling exponents depending on the nature of the critical sector).  In such models, the $\omega \to 0$ and large $N$ limits typically do not commute.  In our theory, in contrast, the `locally critical' sector is replaced by a conventional free
boson.  The marginal Fermi liquid form emerges without tuning, but as with the holographic theories,
the description breaks down at sufficiently low frequency (of order $1/\sqrt{N}$ in the large $N$
model).

For simplicity, we have assumed that the boson bandwidth is small compared to the Fermi energy.  The boson is then reminiscent of an optical phonon that is tuned towards criticality.  The microscopic justification for this assumption can be provided in a multiband system with both localized and itinerant electronic modes;  the bosons  are derived from localized electron degrees of freedom, and strongly affect the  itinerant electrons.   The marginal Fermi liquid behavior that we discuss here would be present up to temperature scales corresponding to the effective `Debye temperature' of the boson.  However above such a scale, the boson behaves effectively as a classical oscillator, and its effect on the fermion modes is also likely to produce deviations from ordinary Fermi liquid behavior.

\acknowledgements{We acknowledge important conversations with S. Chakravarty, A. Chubukov, S. Hartnoll, S. Kivelson, J. McGreevy, M. Mulligan, J. Polchinski, S. Sachdev, E. Silverstein and S. Yaida. SK also thanks the participants of the Simons Symposium on `Quantum Entanglement: From Quantum Matter to String Theory' for providing helpful perspective.  This work was supported in part by the National Science Foundation grant PHY-0756174 (SK) and DOE Office of Basic Energy Sciences, contract DE-AC02-76SF00515(SK and SR), and the Alfred P. Sloan Foundation (SR).
RM is supported by a Gerhard Casper Stanford Graduate Fellowship.
DR is supported by a Morgridge Family Stanford Graduate Fellowship.}

\bibliography{qcmetal}

\end{document}